**Towards the science of living structure: Making and remaking livable cities as part of Urban Informatics**

Bin Jiang[1], Qianxiang Yao[1], Huan Qian[1], and Bisong Hu[2]

[1]LivableCityLAB, Urban Governance and Design Thrust, Hong Kong University of Science and Technology (Guangzhou), China

[2]School of Geography and Environment, Jiangxi Normal University, Nanchang, China

*(Draft: May 2024, Revision: July, August, and September 2024)*

**Abstract**
This chapter investigates the concept of living structure– which is defined as a structural hierarchy that has a recurring pattern of an abundance of small substructures compared to larger ones – and the application of such structures in creating livable cities within urban informatics. By integrating practical, scientific, and artistic innovations, living structures provide a theoretical framework for designing healing environments and understanding urban complexity. We conceptualize spaces through hierarchical transformations, calculating livingness ($L$) as $L = S \times H$, where $S$ is the number of substructures and $H$ is the inherent hierarchy of those substructures. Living structure is governed by the scaling law and Tobler's law, and guided by differentiation and adaptation principles, and it fosters vibrant and engaging spaces that enhance human well-being and a sense of place. Urban informatics, urban planning, and architecture must evolve beyond just understanding and prediction to include purposeful design. This new kind of science integrates the theory of living structure and emphasizes creation and design, thus transforming those disciplines. This chapter looks at environments that have high structural beauty, as defined by the 15 properties that Christopher Alexander proposed, and discusses the application of those properties in architecture, urban informatics, and emerging technologies such as artificial intelligence, with the aim of making built environments more vibrant and conducive to human well-being.

## 1. Introduction

Contemporary fields of architectural and urban design are going through a major transformation, which has been driven by growing acknowledgement of the limitations and failures of modernist approaches. This transformation was highlighted by an open letter (Architectural Education Declares 2019) that over 2500 students and professionals signed, which criticized the lack of sensitivity that current methodologies have regarding environmental and social contexts and called for a new paradigm emphasizing human well-being, sustainability and a harmonious relationship with nature. This institutional critique runs parallel to a grassroots movement called the Architectural Uprising (2014), which supports a return to traditional architectural principles, with an emphasis on local materials, craftsmanship, and vernacular styles. Advocates of the Architectural Uprising argue that such principles are more suitable for the creation of beautiful, livable, and enduring urban environments. They call for designs that resonate with local culture and human experience, rejecting what they believe is the often unsustainable, cold, and impersonal nature of modernist architecture. These calls are supported by the Chernobyl Paradox (Mehaffy and Salingaros 2020), which highlights the long-term consequences of poorly conceived architectural and urban interventions and warns of the dangers of overlooking the interplay between human settlements and their environments.

In the context of urban design, Jacobs (1961) claimed that a lot of modern urban planning was based on nonsensical principles that do not account for cities' organic and intricate nature. Jacobs challenged the mechanistic views of urban spaces, instead calling for cities to be understood as dynamic, living systems. More recent critics of contemporary urban design have included Cuthbert (2007) and Marshall (2012), who argued that urban design practices often lack solid scientific foundations and rely on oversimplified or outdated models that do not capture the complex nature of actual urban environments. Together, such movements and critiques underline the need for architecture and urban design to create a new conceptual framework that embraces sustainability, complexity, and human-centered principles. The present chapter introduces the concept of living structure as a transformative paradigm that attends to this need. Understanding and designing urban environments as living structures makes it possible to create spaces that are functional and also resonant with human experiences and ecological realities.

The science of living structure dramatically changes our understanding of environmental and spatial complexity. A central aspect of this science is the theory of centers (Alexander 2002–2005), which is also known as substructures and is supported by 15 fundamental properties (see Section 2) that define the intrinsic vitality and beauty of a space. This theory is operationalized through innovative methods, such as mathematical models that compute the degree of livingness (Salingaros 1997, Jiang 2015, Jiang and de Rijke 2023), and the mirror-of-the-self experiment (Alexander 2002–2005), which utilizes uses humans and their biometric data obtained from functional near-infrared spectroscopy (fNIRS), eye-tracking, functional magnetic resonance imaging (fMRI), and electroencephalograms (EEG) as measuring instruments with which to gauge the psychological and emotional impact of a space (Sussman 2014, Salingaros and Sussman 2020). Living structure leverages open-access geospatial big data to analyze and enhance urban environments through the integration of such diverse sources as OpenStreetMap, nighttime light imagery, and social media data. This multifaceted approach makes it possible to better understand and create spaces that are not functional, profoundly beautiful, and life-enhancing.

The remainder of this chapter is structured as follows. Section 2 examines the concept of living structure and its 15 properties. Section 3 investigates structural beauty and the vitality of space, as quantified by the formula $L = S \times H$, where $L$ represents livingness, $S$ indicates the number of substructures, and $H$ is their inherent hierarchy. Section 4 presents three illustrative case studies – Alexander's 253 patterns, two pairs of satellite images, and a design example – show the application of the principles of living structure. Section 5 looks at the science of living structure, proposes a shift from a mechanistic to an organic view of space, and outlines a research agenda. Section 6 summarizes the main points and implications.

**2. Living structure and its geometrical and transformation properties**
Christopher Alexander's theory of living structure forms the foundation of the science of living structure, identifying 15 fundamental properties that are universally present in living structures. These properties, which include levels of scale, strong centers, alternating repetition, thick boundaries, and positive space (Figure 1), capture the essence of what makes a structure beautiful and living, and serve as guidelines for designing engaging, coherent, and vibrant spaces. Adherence to these properties can enable a designer to create an environment that fosters a sense of well-being and attachment.

The first of these properties, and the most important, is the levels of scale, which makes sure that a structure has multiple levels of scale, ranging from small details to large overarching forms. This property can be formulated as the scaling law, which is the recurring notion of far more smalls than larges (Jiang 2015). This concept, which is also known as Scale Jump (Alexander 2002–2005), will be discussed further in the next section, with a description of structural beauty or the liveliness of space. The second property is strong centers, which refers to the creation of strong and coherent centers that organize the surrounding elements (or, more precisely, centers or substructures). This property is actually redundant, given that all of the other properties are intended to create strong centers, probably with an exception of the first and the last properties. The third property is thick boundaries, which defines the characteristics of boundaries that distinguish different areas and elements (or substructures in general) within the structure. Next comes alternating repetition, which refers to the use of alternating patterns to create movement and rhythm within the space; this contrasts with the monotonous repletion that is often apparent in modernist buildings and cities. The next property is positive space, which ensures that both the figure and the ground are well-shaped and contribute to the overall form. The sixth property is good shape, which is best understood in a recursive manner, as meaning that a good shape consists of numerous good shapes. According to the recursive definition, this property of good shape aligns with the living structure itself and is therefore considered to be redundant. The seventh property, Local Symmetries, incorporates symmetrical elements to create balance and harmony and mainly refers to symmetries at local levels rather than on the global levels.

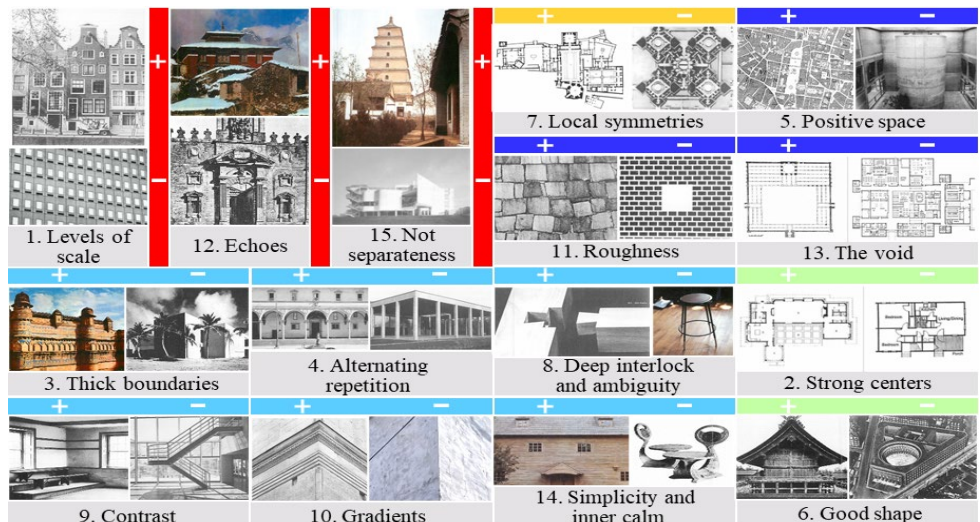

Figure 1: (Color online) The 15 geometrical and transformation properties of living structure
(Note: Each property is represented by positive (+) and negative (-) examples, primarily from Alexander (2002–2005). The examples are placed into five categories, each marked by a different color. Red indicates global properties that span different scales, while the properties in the two blue groups are local. The 'local symmetries' property is highlighted separately because of its unique, pre-existing discovery prior to the set of the 15 properties. 'Strong centers' and 'good shape' are considered to be redundant because they are essentially living structures themselves.)

The eighth property is *deep interlock and ambiguity*, which creates complex relationships between adjacent substructures that enhance the depth and richness of a space. Next is *contrast*, which means using contrast to highlight differences and create visual interest, followed by *gradients*, which means shoring or creating gradual changes in color, form, or texture to create a sense of flow and continuity. The eleventh property is *roughness*; that is, incorporating variations and irregularities in order to avoid monotony and to create a natural appearance. The twelfth property, *echoes*, means repeating substructures to create a sense of coherence and unity at different levels of scale. Property number 13 is *the void*, which refers to the inclusion of open spaces that provide contrast and enhance the surrounding substructures. Next is *simplicity and inner calm*, which means striving for simplicity and calmness in some substructures at the same time as maintaining complexity across various levels of scale. Finally, the fifteenth property is *not separateness*, which ensures that substructures are well integrated and not isolated from one another.

A structure that exhibits these 15 properties to a high degree is considered vibrant and beautiful. However, these properties should be applied holistically, not in isolation. The three types of substructures that need to be considered in any given space are smaller spaces within it, adjacent spaces around it, and the larger space that contains it. Alexander (2002–2005) stated that a space is deemed to be 'living' or 'alive' if it satisfies all three of these criteria, meaning that a space is living when its adjacent spaces, the larger containing space, and the smaller spaces within it are all living. These 15 properties can be further distilled into two fundamental laws: the scaling law (Jiang 2015), which posits that across different scales, there are significantly more small substructures than large ones; and Tobler's law (Tobler 1970), which states that substructures at any level of scale are more or less similar to one another.

## 3. Structural beauty or spatial vitality

The concept of structural beauty – or the livingness of space (*L*) or spatial vitality – is a critical aspect of living structure that underlines the intricate relationship between the number of substructures (*S*) within a space and the hierarchical arrangement of those substructures (*H*). This relationship is expressed mathematically as $L = S \times H$. In order to understand and measure a space's livingness, it is necessary to look at its organic and hierarchical properties. From an organic perspective, space is envisioned in a similar way to biological cells that undergo transformation. Cells divide and form many substructures at various hierarchical levels, and space evolves in the same way. This perspective is in keeping with the principle that a space or subspace has a certain degree of livingness when it embodies far more small substructures than large ones. This recurring notion, which is also known as the levels of scale, hierarchy, or scale jumps, reiterates the essence of complexity within a living structure.

The measure of livingness in a space is tied to its complexity, which follows specific principles and patterns. The two key parameters that define this complexity are the number of substructures (*S*) and hierarchical arrangement (*H*). The former refers to the number of distinct, identifiable subspaces within a larger space, which can range from minute details to larger and more prominent features. When the number of substructures is higher, the livingness of the space is richer. *H* denotes the organization of substructures into different levels or scales. In this context, hierarchy means that smaller substructures are nested within larger ones, which creates a multi-layered structure. As this hierarchy becomes more intricate and well-organized, the livingness increases. Together, the two parameters – *S* and *H* – contribute to the overall livingness (*L*) of a space. The interaction between many small substructures and their hierarchical arrangement creates a dynamic and engaging environment that reflects the principles of living structure.

The notions of scale jumps, hierarchy, levels of scale, and the recurring concept of having far more smalls than larges all describe the same fundamental idea, which is the presence of a complex and multi-level structure in which small elements are embedded within larger ones. This principle can be seen in both natural and human-made (Simon 1962, Bak 1996), ranging from the branching patterns of trees to the arrangement of a traditional urban environment. Figure 2 below illustrates this concept further by comparing two buildings. The image on the left-hand side represents a traditional, organic building that is a landmark at Sun Yat-Sen University in Guangzhou, China, while the image on the right-hand side is a modernist building, Villa Savoye in Poissy, France, designed by Le Corbusier (1923), widely acknowledged as the father of modernist architecture. The traditional building has a livingness of $L(\text{left}) = (3 + 6 + 13 + 51 + 109 + 404) \times 6 = 3516$, while the livingness of the modernist one is $L(\text{right}) = (3 + 14 + 33 + 144) \times 4 = 776$. The marked difference between the two buildings emphasizes how traditional buildings, which have many small and nested substructures, have considerably higher livingness than modernist buildings.

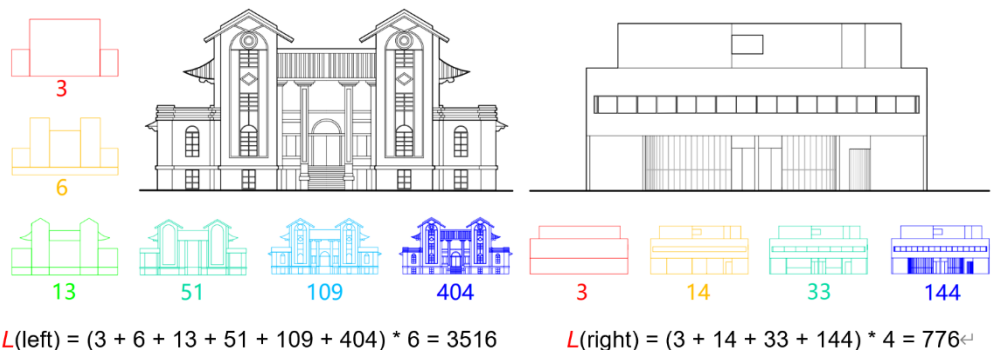

$L(\text{left}) = (3 + 6 + 13 + 51 + 109 + 404) * 6 = 3516$     $L(\text{right}) = (3 + 14 + 33 + 144) * 4 = 776$

Figure 2: (Color online) The building on the left is more living than that on the right

The concept of livingness or vitality has significant implications for architecture and urban planning. Designing spaces that adhere to the principles of living structure will enable architects and planners to create environments that are aesthetically pleasing and also promote well-being and a sense of place (Tuan 1977, Lewicka 2011, Seamon 2018). Rich substructures and their hierarchical organization can change ordinary spaces into vibrant and engaging environments. For instance, a traditional urban setting will often exhibit a high level of livingness due to its intricate street patterns, layered public spaces and diverse building forms. By contrast, a modernist design that emphasizes uniformity and simplicity may lack such complexity, which will result in less engaging environments. The livingness of a space reflects its complexity and structural beauty. Focusing on the number of substructures and their hierarchical arrangement enables us to better understand and enhance the livingness of built environments. Such an approach aligns with scientific principles and resonates with humans' innate preference for complex and richly structured spaces.

The two fundamental design principles for creating living structures – differentiation and adaptation – align with the two above-mentioned core laws of living structures. Differentiation is the process of transforming space through cell-like division, which generates multiple substructures across various scales, ensuring that each structure is unique and contributes to the richness of the overall form. Adaptation makes sure that all generated substructures harmonize with each other; while substructures can vary in terms of size and form, they require a level of coherence to ensure that each element relates to the structures surrounding it in such a way that preserves overall unity. While many other properties focus on achieving adaptation, the property of levels of scale mainly supports differentiation. These two principles – differentiation and adaptation – have been distilled from the 15 fundamental properties of living structure, which reflects the vital role they play in the production of spaces that are at once complex and coherent.

### 4. Illustrative case studies
This section presents three case studies that show the principles of living structure in different contexts. Each of the case studies applies hierarchical organization and the concept of "far more smalls than larges" to show the structural beauty and complexity of the subjects. The first case study looks at Alexander's 253 patterns and highlights their coherent network of substructures. The second study compares nighttime satellite images in order to assess livingness in both urban and rural areas. In the third case study, we detail the transformation of a classroom, highlighting hierarchical differentiation in creating engaging spaces. Together, these case studies showcase impact of practical applications of living structure theory in urban planning and design on various environments.

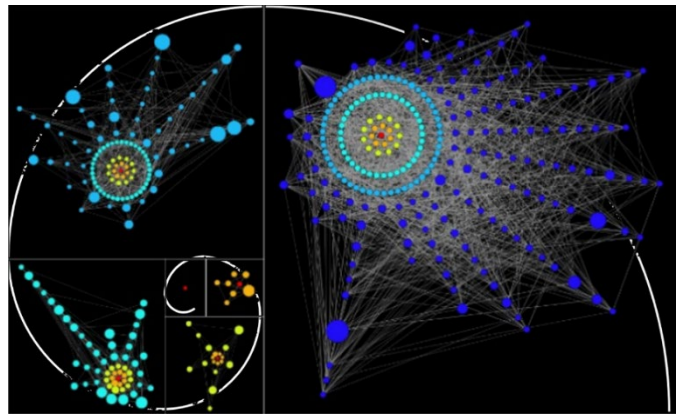

Figure 3: The notion of far more smalls than larges recurs multiple times

### 4.1 The 253 patterns
The concept of living structure can be traced back to the 253 patterns, defined by Alexander et al. (1977), that address a range of problems and solutions. Each pattern is connected to both larger and smaller patterns, which creates a network like that of substructures. This case study investigates how these patterns form a coherent whole and computes their degrees of order or complexity. We organized the 253 patterns into six hierarchical levels, ranging from the largest (world government) to the smallest (daily life objects), which reveals a clear scaling hierarchy with far more small patterns than large ones. We then constructed a topological network based on the initial definitions of pattern associations. In contrast to previous studies, we arranged the patterns hierarchically and only allowed links within the same level or between adjacent levels, to ensure a clear hierarchical structure.

In this case study, we treated patterns as substructures that form a living structure. This is shown in Figure 3 in a spiral-shaped layout, which starts with the largest patterns and evolves to the smallest patterns, in the lowest hierarchical level, which exhibit the highest degrees of order or structural beauty; this reflects the principle that stronger differentiation leads to higher order. Each new generation of patterns emerges from the previous generation, inheriting and increasing the degree of order or complexity. This stepwise emergence can be seen in Figure 3, where each successive level is more orderly or more complex than the previous one. The 253-node network shows that the degree of order correlates with the recurring notion of far more smalls than larges, which makes the network visually appealing because it embodies structural beauty independent of color.

This case study shows how the 253 patterns form a coherent living structure. The hierarchical arrangement and the emergence of patterns at different scales reflect organized complexity. Such an approach quantifies the degree of order and also demonstrates how adaptation and differentiation principles guide the creation of living structures. This process mirrors natural human settlement patterns and emphasizes the importance of maintaining such principles in modern urban planning to avoid the creation of nonliving structures.

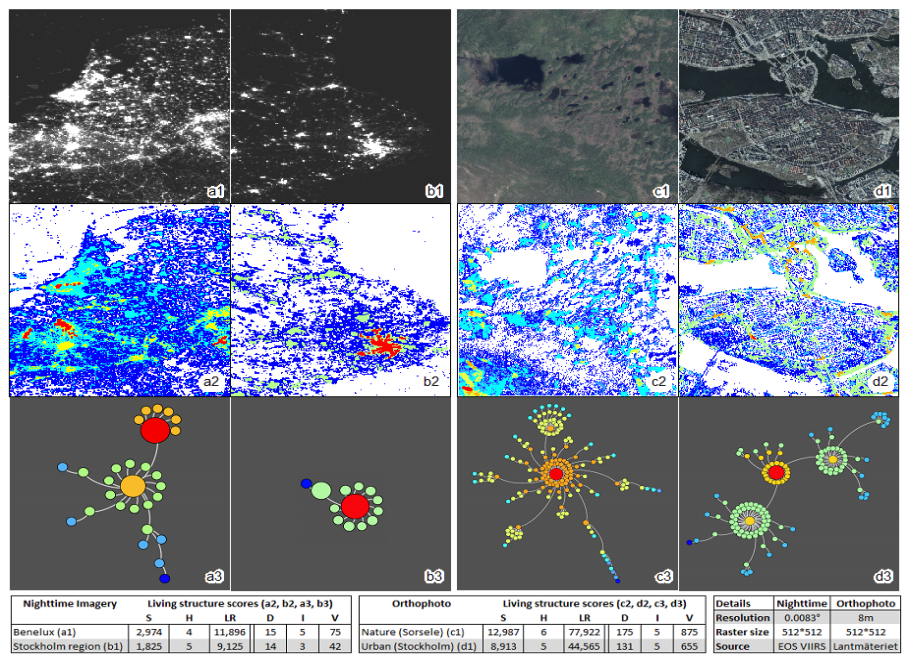

Figure 4: (Color online) The recursive approach applied to the four georeferenced images
(Note: The source images (a1– d1); their corresponding living structures, both decomposable and undecomposable (a2– d2); and the decomposable living structures in graphical format (a3– d3). The living structures shown in the second row (a2– d2) have numerous first-iteration substructures (in blue), a few last-iteration substructures (in red), and some between the first- and last-iteration substructures (in colors on the spectrum between blue and red). Some of the substructures are nested to each other, with different colors showing the nested relationship, as shown in the second and third rows.)

## 4.2 The two pairs of satellite images

The second case study involved comparing nighttime images of the Benelux (Belgium, the Netherlands, and Luxembourg) region with the Stockholm region (Figure 4, Panels a1 and b1). Because Benelux is more densely populated than Stockholm, it features a higher number of small human settlements, which translates to greater livingness, as reflected in the livingness scores (*LR* and *V*) and visualized in Panels a2, b2, a3, and b3 of Figure 4. The high density of small settlements in the Benelux region (Panel a1) creates complex, vibrant living structure, as can be seen in Panels a2 and a3 in the many first-iteration substructures (shown in blue), several intermediate substructures (represented by various colors), and a few last-iteration substructures (red). By contrast, the Stockholm region (Panel b1) has fewer small settlements and exhibits lower livingness scores. Panels b2 and b3 display a less intricate hierarchical structure with fewer substructures.

The second pair of panels compares a satellite image of the Swedish countryside with the center of Stockholm city (Figure 4, Panels c1 and d1). While the two images have the same physical size and resolution, they capture different geographic features. The Swedish countryside (Panel c1) was expected to exhibit higher livingness because of its inherent complexity and variability, and this expectation was confirmed by the higher livingness scores (shown in Panels c2 and c3), displaying a rich structure with many first-iteration substructures (blue), intermediate substructures (various colors), and a few last-iteration substructures (red). The urban scene of Stockholm (Panel d1) shows lower livingness scores due to its structured and less complex layout, while Panels d2 and d3 reveal a simpler hierarchical structure with fewer substructures. The recursive decomposition of these images presents numerous first-iteration substructures (blue), a small number of last-iteration substructures (red), and intermediate substructures in various colors. Rows 2 and 3 of Figure 4 depict these nested relationships, highlighting each scene's hierarchical complexity.

The recursive approach (LR and V) captures the livingness of georeferenced images, showing higher livingness scores for densely populated and natural scenes than for less populated and urban areas. This method delivers a robust framework with which to quantify and visualize the livingness of different environments, highlighting the intricate structures that contribute to a region's overall sense of liveliness. The recursive approach, in revealing these underlying patterns, generates valuable insights into the spatial organization and complexity of certain geographic areas.

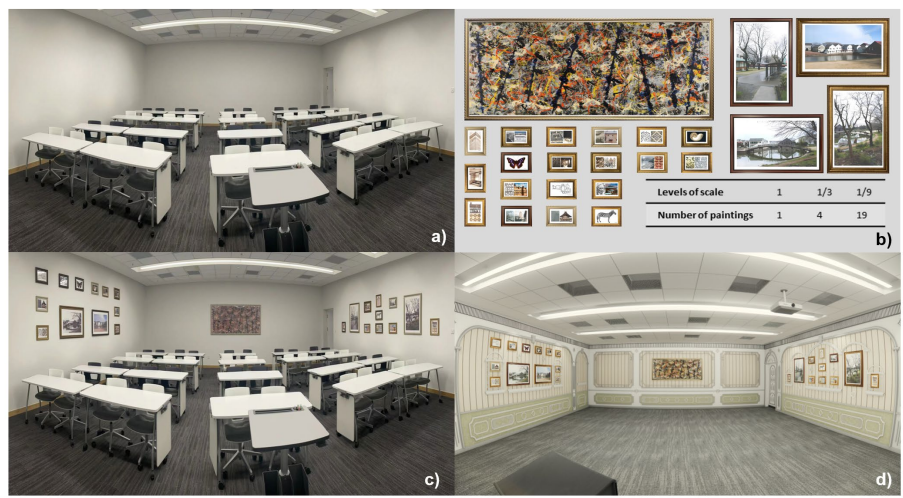

Figure 5: (Color online) A small design process using the levels of scale, or the 15 transformation properties

### 4.3 Renovation of a classroom
Figure 5 shows the transformation of a classroom – through a careful design process that involves a thoughtful approach to how the space evolves and differentiate – from a plain and almost empty space into a vibrant and lively room. Initially, the room was basic and only contained essential furniture (Panel a). The walls are bare, and the overall feeling is sterile and uninviting. This minimal setup provides a 'blank canvas' that highlights the potential for enhancement and the importance of thoughtful design to create a more engaging environment. Panel b shows the planning phase, in which different levels of scale were considered, including decisions regarding the number and size of the pictures to be hung. The table in Panel b shows three levels of scale – 1 (large), 1/3 (medium), and 1/9 (small) – with a corresponding number of paintings at each scale (one large, four medium, and 19 small). This careful consideration means that the elements are distributed in a balanced and aesthetically pleasing manner.

This step is crucial because it lays the foundations for a harmonious design that emphasizes proportion and variety in visual interest.

Panel c shows the room starting to come to life. Pictures with varying sizes are placed strategically on the walls following the planned levels of scale, creating an engaging environment. The room feels warmer and more inviting than it did initially, while the artwork adds character and interest. The intentional placement of each picture shows how layering and including diverse elements can make a space more dynamic and stimulating. Panel d shows the room reaching its fully developed state. The design includes the pictures on the walls and integrates substructures that consider both the figure (the pictures) and the ground (the walls and the overall space). The room feels cohesive and harmonious, with the various substructures acting in conjunction to create a stimulating and lively environment. This final transformation shows the culmination of the design process, with all of the substructures contributing to a unified aesthetic that enhances the feel and function of the space. This small design process shows how paying careful attention to the arrangement and scale of the various substructures can turn a plain room into a vibrant and welcoming space. This approach reinforces the importance of design when creating functional and enjoyable environments.

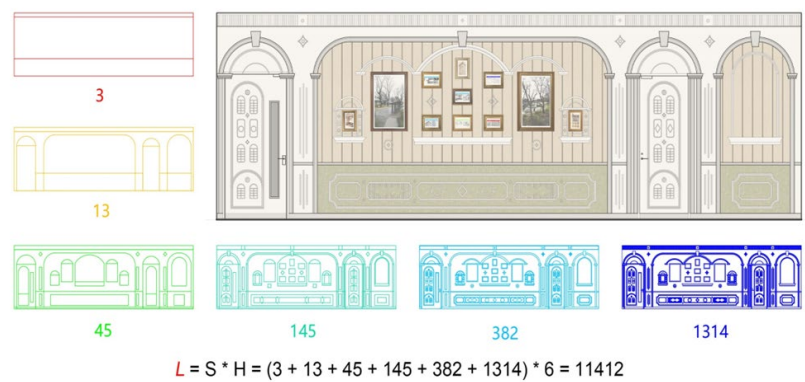

Figure 6: Substructures defined at different levels of scale for a wall of the classroom

Figure 6 shows a wall and its detailed substructures, particularly the inherent hierarchy that contributes to the wall being classified as a living structure with a very high degree of livingness. The wall has multiple layers, with each having its own level of complexity and scale. The substructures are organized hierarchically, ranging from the largest elements (labeled with the number 3) to progressively smaller and more detailed components (ending with the number 1314). This hierarchy reflects how natural organisms are built, with larger systems encompassing smaller, detailed sub-systems. Each substructure is interconnected with others to form a coherent whole, ensuring that changes in one part of the structure influence all the other parts and creating a responsive and dynamic system. For example, the largest substructures provide the main framework, while the smaller elements add detail and refinement, which enhances the overall aesthetic and functional qualities.

Using different levels of scale, with respective numbers of substructures—3, 13, 45, 145, 382, and 1314—demonstrates a thoughtful consideration of balance and proportion. Larger substructures provide the foundation, offering structure and stability, while smaller substructures add depth and intricacy, making the wall visually engaging and rich in detail, much like a living organism that can be observed at micro and macro levels. In addition to the levels of scale, the design incorporates multiple other essential characteristics of a living structure such as thick boundaries, local symmetry, and alternating repetition. This alternating repetition can be seen in the consistent use of panels, arches, and decorative motifs all across the wall. The wall's design process mimics organic growth, starting with the basic framework and progressively adding layers of detail. This approach is similar to the development of an embryo, with the entire structure giving rise to its parts through differentiation. The formula $L = S \times H = (3 + 13 + 45 + 145 + 382 + 1314) \times 6 = 11{,}412$ shows the cumulative complexity or livingness and the holistic nature of the design, with each part contributing to the living whole. By considering these substructures, the wall shifts from being a mere static structure to a living entity that is rich in detail, coherence, and organic complexity. This approach to design ensures that the space it inhabits is harmonious, engaging, and dynamic, embodying the principles of living structures found in nature.

## 5. Further research on the science of living structure

The science of living structure represents a shift from a mechanistic view of space to an organic one. The mechanistic view (Descartes 1644, 1983) perceives space as a compartmentalized and static entity that is focused on function and efficiency, which often leads to rigid and impersonal urban designs that are lacking in coherence and fail to engage users emotionally. The organic view (Bohm 1983, Whitehead 1929) considers space as a living system that is adaptive, interconnected, and flexible, emphasizing complexity, structural beauty, and human experience, with all elements interacting harmoniously with their surroundings. This shift is embodied in Alexander's 15 properties of living structure, all of which reflect the interconnectedness found in nature. These properties are distilled into two key design principles: differentiation (the generation of distinct but related substructures) and adaptation, whereby the substructures fit together to create a unified whole. This organic approach helps create urban environments that are functional, aesthetically pleasing, and vibrant, and resonate with human experience.

Within the science of living structures, a critical research agenda is to leverage geospatial big data to reveal and analyze patterns of living structures in urban and architectural settings, which can enhance our understanding of human mobility, spatial dynamics, and the temporal evolution of urban areas. Advanced analytical techniques and computational models inspired by the notion of living structure enable researchers to reveal the underlying hierarchies and substructures that contribute to the livingness of a space. Such knowledge can inform urban planning and policy decisions, making cities more livable, sustainable, and resilient. Integrating geospatial big data, such as OpenStreetMap data, social media data, and nighttime imagery, makes it possible to study the dynamics of living structures and develop insights into how human activities shape and are shaped by their environments.

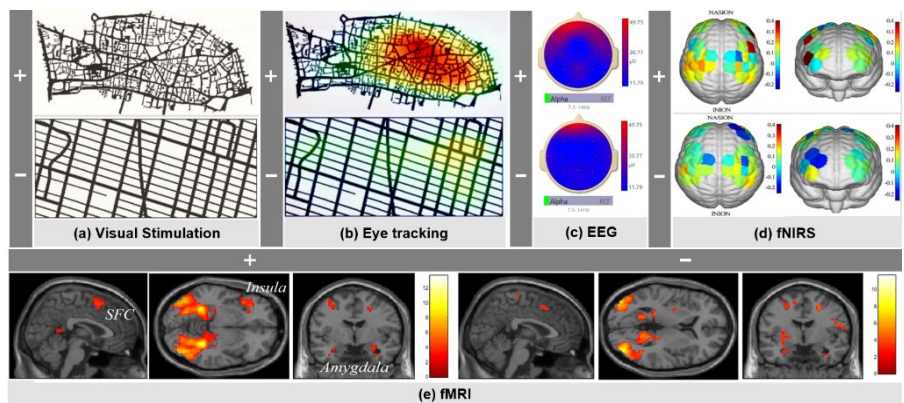

Figure 7: (Color online) Multimodal assessment of neurological responses to structural beauty
(Note: (a) Two urban grid structures with high (+) and low (-) livingness, (b) Eye-tracking heat maps showing gaze patterns, (c) EEG: Brainwave analysis highlighting emotional and cognitive responses, (d) fNIRS: Neural activation maps from visual stimuli, and (e) fMRI: Greater activation for complex (+) forms.)

The concept of structural beauty extends beyond urban planning into a few other fields, including computer vision and image understanding. Applying the principles of living structure makes it possible to develop algorithms that assess the structural beauty of images, which can contribute in such areas as art, architectural design, and artificial intelligence (Jiang and Huang 2021, Jiang and de Rijke 2023). This interdisciplinary approach bridges the gap between functional and aesthetic considerations, fostering environments that are both visually appealing and conducive to human well-being both physical and emotional. As various case studies have shown, the recursive approach to computing the structural beauty of images reveals that images that have higher levels of structural hierarchy and substructures are perceived as more beautiful than those with lower levels. Applying the same methodology to urban landscapes can help identify and enhance the livingness of cityscapes through improved design practices.

There is neuroscientific evidence that living structures have positive effects on human perception and cognition (e.g., Ishizu and Zeki 2011, Kawabata and Zeki 2004). Studies that utilize eye-tracking technology, fMRI, EEG, and fNIRS could show that environments with high structural beauty can reduce stress, enhance cognitive performance, and improve overall mental health (Figure 7 for an illustration). Such studies underline the importance of integrating living structures into urban design, which highlights the potential of this approach to transform modern cities into engaging and vibrant spaces that promote human flourishing. Urban planners and architects who create environments that align with the principles of living structure can help develop a deeper connection between individuals and their surroundings, which can enhance the quality of life in urban areas.

## 6. Conclusion

The science of living structure represents a burgeoning discipline that transcends conventional mechanical frameworks and embraces an organic view of space that helps us understand and enhance the built environment. This approach stresses the intrinsic order and hierarchical structure underlying all spatial forms, marked by many more small substructures than large ones and repeating patterns of similar substructures on various scales. This perspective leverages principles of adaptation and differentiation with the intention of bridging the gap between art science and art to create a holistic framework that acknowledges beauty as an objective property that is rooted in the hierarchical and mathematical organization of space. This science is built on Tobler's law and the scaling laws, which collectively posit that a living structure needs to have many more small elements than large ones, and that the elements must generally be similar within their levels. This approach is organic and recursive and exposes the deep order that exists in spaces, patterns and phenomena that exist in nature and cultivating environments that are in close harmony with the human psyche.

By unlocking the foundational principles of living structure and its applications, this chapter has demonstrated how this field of science can potentially revolutionize architecture and urban planning, as well as urban informatics. Examining structural beauty through the lens of living structure can help transform urban environments in a way that fosters a greater sense of community, environmental sustainability, and well-being. The case studies that we have presented show the ability of the recursive approach to measure, differentiate and enhance the livingness of spaces. Integrating geospatial big data gives us a greater appreciation of space, time, and human mobility and paves the way for a comprehensive sustainable urban planning framework. Ongoing research will investigate living structure's implications in sustainable urban planning, geospatial analysis, and neuroscientific studies of human responses to spatial configurations, among other domains. By recognizing the integral role that human experience and consciousness play in the built environment, the science of living structure opens the door for an increasingly integrated and enriched approach to designing spaces that will promote human well-being and a place identity. This emerging field has the potential to unify artistic and scientific principles, ultimately creating environments that are not only functional, but also deeply fulfilling and inherently beautiful.

**Acknowledgements:**
XXXXXX